# Computable Phenotypes for Post-acute sequelae of SARS-CoV-2: A National COVID Cohort Collaborative Analysis

Sarah Pungitore, MS[1], Toluwanimi Olorunnisola[2], Jarrod Mosier, MD[3], Vignesh Subbian, PhD[2], and the N3C Consortium*

[1]Program in Applied Mathematics, The University of Arizona, Tucson, AZ; [2]College of Engineering, The University of Arizona, Tucson, AZ; [3]College of Medicine - Tucson, The University of Arizona, Tucson, AZ;
*Consortial contributors are in the process of being documented

**Abstract**
*Post-acute sequelae of SARS-CoV-2 (PASC) is an increasingly recognized yet incompletely understood public health concern. Several studies have examined various ways to phenotype PASC to better characterize this heterogeneous condition. However, many gaps in PASC phenotyping research exist, including a lack of the following: 1) standardized definitions for PASC based on symptomatology; 2) generalizable and reproducible phenotyping heuristics and meta-heuristics; and 3) phenotypes based on both COVID-19 severity and symptom duration. In this study, we defined computable phenotypes (or heuristics) and meta-heuristics for PASC phenotypes based on COVID-19 severity and symptom duration. We also developed a symptom profile for PASC based on a common data standard. We identified four phenotypes based on COVID-19 severity (mild vs. moderate/severe) and duration of PASC symptoms (subacute vs. chronic). The symptoms groups with the highest frequency among phenotypes were cardiovascular and neuropsychiatric with each phenotype characterized by a different set of symptoms.*

**Introduction**
The COVID-19 pandemic has had unprecedented effects on healthcare systems worldwide. In the United States alone, there have been an estimated 5.4 million hospitalizations and 1.05 million deaths since August 2020.[1] Much research has been devoted to developing treatment regimes, understanding systemic effects of the virus, and analyzing the public health impact of the pandemic.[2] However, after an acute infection, patients are at risk of post-acute sequelae of SARS-CoV-2 (PASC).[3] PASC (also known as Long COVID) is currently defined as the persistence of COVID-19 symptoms or complications from 3 to 4 weeks following the onset of symptoms from SARS-CoV-2 infection.[4] It is estimated at least 10% of all individuals infected with SARS-CoV-2 and up to 70% of those hospitalized for COVID-19 will develop PASC.[5] Commonly reported symptoms include fatigue, cognitive impairment, difficulty in breathing, and general pain.[6] These symptoms can persist for over six months.[6]

Despite the prevalence of PASC in survivors of COVID-19, there remain many challenges in understanding the disease. One challenge is the lack of a formal, comprehensive clinical definition for PASC. For example, over 200 different symptoms have been associated with PASC across different studies.[7] Additionally, even after an individual is clinically diagnosed, it is difficult to accurately identify positive cases from the ICD-10 diagnostic code U09.9 alone. While considered the primary diagnosis code for PASC, it was only made available in October 2021 and likely does not capture all cases because of the heterogeneity of the disease.[8] Finally, the underlying causes of PASC, which may range from injury to one or more organs or chronic reservoirs of SARS-CoV-2 in specific tissues, have not yet been determined.[3]

Several studies have examined PASC phenotypes to better classify, understand, and treat different disease subgroups. Studies that evaluated phenotypes based on symptomatology found symptom clusters for central sensitization symptoms,[9] cardiovascular symptoms,[10] and respiratory symptoms,[11] along with clusters for general physiologic abnormalities.[12] Other studies developed phenotypes based on age and/or gender,[8,13] symptom duration,[13] and COVID-19 severity.[14] However, these studies leave several gaps in knowledge. First, there is currently no standardized definition for PASC based on symptomatology. Second, there is a lack of computable phenotypes, or heuristics (reproducible and interpretable disease concepts) and meta-heuristics (rules and definitions that structure the development of the computable phenotypes) that can be generalized to multiple data sources. Finally, no studies to our knowledge have jointly evaluated phenotypes based on COVID-19 severity and duration of PASC symptoms.

Therefore, the primary objective of this study was to standardize PASC phenotyping methods by defining computable phenotypes and meta-heuristics for PASC based on COVID-19 severity and symptom duration. This reproducible framework can be applied to various clinical settings to identify and phenotype PASC patients without the ICD-10 code U09.9.

**Methods**
*Data Use and Ethics*
Data use was approved by the National COVID-19 Cohort Collaborative (DUR #RP-AB8693) and the University of Arizona (IRB #1907780973).

*Data Sources and Study Sample*
The data for this study were collected from the National COVID-19 Cohort Collaborative (N3C). N3C is a nationwide effort formed to further COVID-19 research by collecting and harmonizing relevant electronic health record (EHR) data from multiple clinical sites.[15] N3C encompasses an entire data collection and analytics pipeline where EHR data is sourced from participating sites, encoded using a common data model, and accessed exclusively through an analytics platform.[15] The data source for this study was a limited data set provided through the N3C enclave. All data in the N3C enclave are organized and standardized using the Observational Health Data Sciences and Informatics (OHDSI) Observational Medical Outcomes Partnership (OMOP) Common Data Model (CDM),[16] which is required by N3C to facilitate research. Unlike de-identified data, the limited data set contains information on patient dates of service and zip codes.[17] For each visit, timestamped EHR data such as demographics, vital signs and other physical measurements, symptoms, procedures, lab results, medications, and medical conditions were recorded for each patient.[17] Patients were excluded from the analysis if they did not have at least one record of a positive SARS-CoV-2 test based on the SARS-CoV-2 polymerase chain reaction tests concept set (N3C Codeset ID: 651620200) or if they were not $\geq$18 years at the time of the index COVID-19 visit.

*PASC Symptom Concepts*
We identified symptoms relevant for PASC to facilitate phenotype development using PASCLex, a previously developed PASC symptom lexicon.[18] PASCLex symptoms with a frequency of at least 1% were selected for symptom concepts. Additional concepts were identified by searching relevant literature for PASC phenotyping. Concept sets for each symptom were created using ATLAS, an OHDSI-specific tool for interfacing with the OMOP CDM.[19] Expert clinician review was used to determine which symptom-related concepts were appropriate for identifying PASC. Symptoms were clustered into physiological categories (cardiopulmonary, neuropsychiatric, gastrointestinal, endocrine, renal, dermatologic, and systemic).[4,8] The concept set listing all symptom concepts can be accessed in N3C (N3C Codeset ID: 551255549). Symptom concepts were sourced from the OMOP *note* and *condition_occurrence* tables for analysis.[20]

*Meta-heuristics for PASC Phenotypes*
We established the following meta-heuristics i.e., rules to structure the development of PASC phenotypes:
1) Define relevant timepoints (Figure 1):
    a. COVID-19 index date: the date of first infection based on the SARS-CoV-2 PCR tests concept set (N3C Codeset ID: 651620200).
    b. Second COVID-19 index date: the date of second infection, if it exists, based on SARS-CoV-2 PCR tests concept set (N3C Codeset ID: 651620200). This date must be at least 90 days following the COVID-19 index date.
    c. Start of subacute PASC: the date 4-weeks (28 days) following the COVID-19 index date after which individuals can start to be diagnosed with subacute PASC.[4]
    d. Start of chronic PASC: the date 12-weeks (84 days) following the COVID-19 index date after which individuals can start to be diagnosed with chronic PASC.[4]
2) Define relevant timeframes (Figure 1):
    a. Pre-COVID-19: the time period up to, but not including, the COVID-19 index date.
    b. Acute COVID-19: the time period from the COVID-19 index date up to, but not including, the start of subacute PASC.
    c. Subacute PASC-susceptible period: time from start of subacute PASC up to, but not including, the start of chronic PASC.
    d. Chronic PASC-susceptible period: time from end of subacute PASC.
3) Identify patients with a second infection by presence of a second COVID-19 index date. When a second infection overlaps with the PASC-susceptible periods, the patient is no longer considered PASC-susceptible.

4) Identify PASC symptoms present in both the PASC-susceptible periods and the pre-COVID-19 periods, using OMOP concept IDs for the selected symptom concepts.
5) Discard any symptoms present for each patient during the pre-COVID-19 period that were also present in the PASC-susceptible periods to avoid associating pre-existing conditions with PASC.
6) Identify which symptoms were associated with subacute or chronic PASC based on whether the symptom was present in the subacute or chronic PASC-susceptible period.
7) Classify the index COVID-19 into mild or moderate/severe. The National Institute of Health (NIH) clinical spectrum of COVID-19 includes five stages of illness from asymptomatic to critical illness.[21] However, due to the variations in N3C site reporting, measurements related to these criteria were not present for every patient. Therefore, we provide a more general framework based on visit type and visit duration, while using the NIH clinical measurement guidelines for classification if visit information was unknown, but measurements were still recorded. Visits that could not be classified either according to visit type, visit duration, or clinical measurements were classified as unknown.
   a. Mild COVID-19:
      i. The index visit is not associated with an inpatient visit or an emergency department (ED) visit followed by admission. If the index visit is to the ED without information on admission, the duration of the ED visit should be < 24 hours **or**
      ii. The first respiratory rate recorded does not exceed 30 breaths/min or the first recorded $SpO_2$ value was 94% or higher.
   b. Moderate/Severe COVID-19:
      i. The index visit is an inpatient visit, an ED visit followed by an inpatient visit, or an ED visit of ≥ 24 hours when information on admission is unknown **or**
      ii. The first respiratory rate recorded exceeds 30 breaths/min or the first recorded $SpO_2$ value was below 94%.

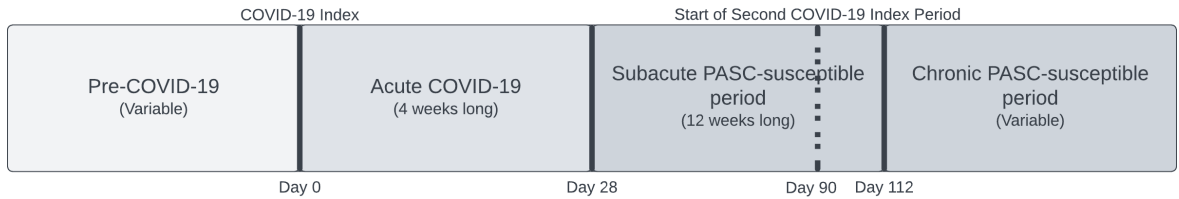

**Figure 1.** Overview of timeframe meta-heuristics used for phenotyping. Dashed line represents the start of the second COVID-19 index, which may not be present for all patients.

*Computable Phenotypes for PASC*
We developed computable phenotypes (i.e., heuristics) to identify patients with PASC based on the meta-heuristics and the symptom profile for PASC (Table 1).

**Table 1.** Computable PASC phenotypes based on symptom duration and COVID-19 severity.

| Phenotype | Criteria (Presence of one or more PASC symptom concepts in the:) |
| --- | --- |
| Subacute PASC, mild COVID-19 | Subacute PASC-susceptible period following mild index COVID-19 |
| Subacute PASC, moderate/severe COVID-19 | Subacute PASC-susceptible period following moderate/severe index COVID-19 |
| Chronic PASC, mild COVID-19 | Chronic PASC-susceptible period following mild index COVID-19 |
| Chronic PASC, moderate/severe COVID-19 | Chronic PASC-susceptible period following moderate/severe index COVID-19 |

*Characterization of PASC Phenotypes*
Using the phenotyping heuristics, we characterized patients based on age, gender, and the number and type of symptoms. Age (year) and gender (male, female, or unknown) for each patient were determined from demographic data associated with the index COVID-19 date. We also compared the phenotyped patients against those clinically diagnosed with PASC. Clinically diagnosed patients were identified by presence of the ICD-10 code U09.9 at least four weeks following the COVID-19 index date.

## Results
*Study Sample*
After applying the inclusion and exclusion criteria, there were 4,176,352 patients. After phenotyping, we identified 907,391 patients with PASC. Of the original 4,176,352 patients, 20,292 were clinically diagnosed with PASC while 15,091 of the 907,391 phenotyped patients were also clinically diagnosed with PASC. Therefore, the incidence of PASC in the entire cohort increased from 0.5% based on clinical diagnosis to 21.7% based on phenotyping. Baseline characteristics for clinically diagnosed and phenotyped PASC patients are presented in Table 1. Among those clinically diagnosed with PASC, a higher percentage of patients had more than one SARS-CoV-2 infection when compared to the phenotyped patients. Additionally, on average, the phenotyped patients were younger than clinically diagnosed patients.

**Table 2.** Baseline characteristics for phenotyped and clinically diagnosed PASC patients. Phenotyped PASC patients were those identified from the computable phenotypes outlined in Table 1. Clinically diagnosed patients were identified by presence of the ICD-10 code U09.9 at least four weeks following the COVID-19 index date. The number of patients may not sum to the total due to missing values.

|  | **Phenotyped PASC Patients (n = 907,391)** | **Clinically Diagnosed PASC Patients (n = 20,292)** |
|---|---|---|
| Age at COVID-19 index, mean (SD) years | 49.2 (18.3) | 54.0 (15.9) |
| Gender, n (%) |  |  |
|   Male | 334,711 (36.9) | 7,220 (35.6) |
|   Female | 572,552 (63.1) | 13,071 (64.4) |
| COVID-19 Severity, n (%) |  |  |
|   Mild | 756,850 (83.4) | 17,270 (85.1) |
|   Moderate/Severe | 53,166 (5.9) | 2,120 (10.4) |
|   Unknown | 97,375 (10.7) | 902 (4.4) |
| Number of SARS-CoV-2 Infections |  |  |
|   1 | 856,381 (94.4) | 18,246 (89.9) |
|   2 | 49,844 (5.5) | 1,959 (9.7) |
|   3+ | 1,166 (0.1) | 87 (0.4) |

*PASC Symptoms*
After identifying symptoms from PASCLex and other literature,[9,10,13,14] we identified 6,569 concepts that mapped to 151 symptoms. After expert review, there were 405 concepts that mapped to 74 symptoms. The symptoms were grouped into seven physiological groups (Table 3). Of these 74 symptoms, 72 were present among the 4,176,352 patients in the N3C study cohort with 69 present in patients not clinically diagnosed with PASC and 72 present in patients clinically diagnosed with PASC.

**Table 3:** PASC symptoms selected by expert review from PASCLex.[18]

| **Physiologic Group** | **Symptoms in N3C Cohort (n = 72)**[*] |
|---|---|
| **Cardiopulmonary** | abnormal breathing, bradycardia, bronchospasm, chest discomfort, chest pain, cough, dyspnea, nasal congestion, palpitations, respiratory syndrome, tachycardia, wheezing |
| **Neuropsychiatric** | abnormal vision, aggressive behavior, agitation, altered mental status, anger, anosmia, anxiety, aphasia, clouded consciousness, confusion, dizziness, fatigue, flat affect, hallucinations, headache, impaired cognition, insomnia, irritable, memory impairment, mood swings, pain, panic, paresthesia, poor concentration, restless legs, seizure, self- |
| | harm, sleep apnea, sleep disorder, somnolence, suicidal, taste/smell issues, tinnitus, tremor |
| **Gastrointestinal** | abdominal bloating, bloody stools, constipation, diarrhea, gastroesophageal reflux, gastrointestinal hemorrhage, incontinence feces, loss of appetite, nausea/vomiting, restlessness, suffering, ulceration |
| **Dermatologic** | dermatitis, erythema, itching, rash, urticaria |
| **Endocrine** | dysmenorrhea, menstruation abnormalities, vaginal bleeding |
| **Renal** | abnormal uterine bleeding |
| **Systemic** | arthralgia, fever, lymphadenopathy, myalgia, weakness |

*The two symptoms missing from the N3C cohort were weight loss (systemic) and stiffness (systemic).

*Characterization of PASC Phenotypes*

After applying the meta-heuristics, there were 756,850 patients with mild index COVID-19, 53,166 patients with moderate/severe index COVID-19, and 97,375 patients with an unknown index COVID-19 (Table 2). Of the patients with mild index COVID-19, 278,556 had symptoms recorded in the subacute PASC timeframe and 613,072 had symptoms recorded during the chronic PASC timeframe. Of the patients with moderate/severe index COVID-19, 27,692 had symptoms recorded in the subacute PASC timeframe and 37,983 had symptoms recorded during the chronic PASC timeframe. Table 4 provides baseline characteristics for each phenotype. Overall, a higher proportion of females with mild index COVID-19 had PASC symptoms when compared to males. Patients with moderate/severe index COVID-19 were also older on average than those with mild index COVID-19. The average number of symptoms was similar across COVID-19 severity and symptom duration.

**Table 4.** Baseline characteristics of the phenotyped patients based on the phenotypes outlined in Table 1. The number of patients may not sum to the total due to missing values.

|  | Subacute PASC | | Chronic PASC | |
|---|---|---|---|---|
|  | **Mild COVID-19 (n = 278,556)** | **Moderate/Severe COVID-19 (n = 27,692)** | **Mild COVID-19 (n = 613,072)** | **Moderate/Severe COVID-19 (n = 37,983)** |
| Age at COVID-19 index, mean (SD) years | 49.7 (18.3) | 60.6 (17.7) | 48.2 (17.9) | 59.0 (17.8) |
| Gender, n (%) | | | | |
| Male | 100,075 (35.9) | 13,481 (48.7) | 218,652 (35.7) | 17,480 (46.0) |
| Female | 178,438 (64.1) | 14,209 (51.3) | 394,339 (64.3) | 20,501 (54.0) |
| Number of PASC symptoms, mean (SD) | 1.8 (2.2) | 2.1 (3.0) | 2.2 (2.6) | 2.4 (3.0) |

The symptom groups with the highest symptom frequency were neuropsychiatric and cardiopulmonary, regardless of index COVID-19 severity or symptom duration (Figure 2). However, unlike the other phenotypes, patients with moderate/severe index COVID-19 and subacute PASC symptoms had a higher frequency of cardiopulmonary symptoms than neuropsychiatric symptoms. The most frequent individual symptoms varied between phenotypes (Figure 3). For patients with mild and moderate/severe index COVID-19, dyspnea was the most frequently reported symptom during the subacute PASC timeframe. Additionally, certain neuropsychiatric symptoms, such as anxiety and pain, were more frequent in chronic PASC than subacute PASC regardless of the severity of the index COVID-19. Certain gastrointestinal symptoms, such as diarrhea and constipation, were most frequently present in patients with moderate/severe index COVID-19 when compared to those with mild index COVID-19 while certain cardiopulmonary symptoms, such as chest pain and palpitations, were most frequently present in patients with mild index COVID-19 when compared to those with moderate/severe COVID-19.

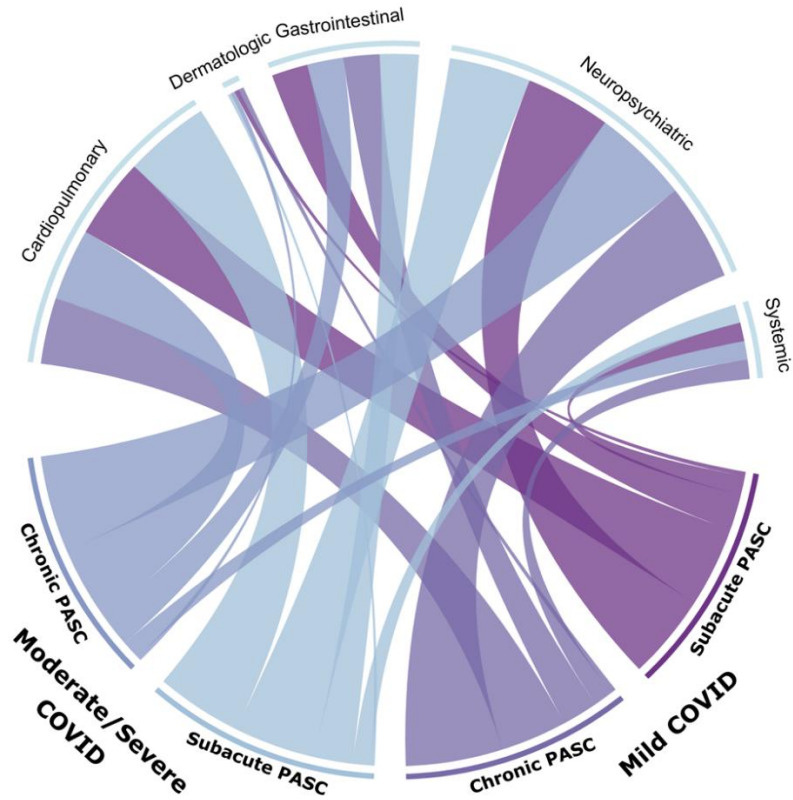

**Figure 2.** Chord plot illustrating distribution of symptom groups (top half of the circle) across the four phenotypes (bottom half of the circle). Links represent frequency of patients with symptoms in each symptom group. Links are ordered counterclockwise from highest to lowest frequency in each phenotype. Renal and endocrine symptom groups were removed due to low symptom counts.

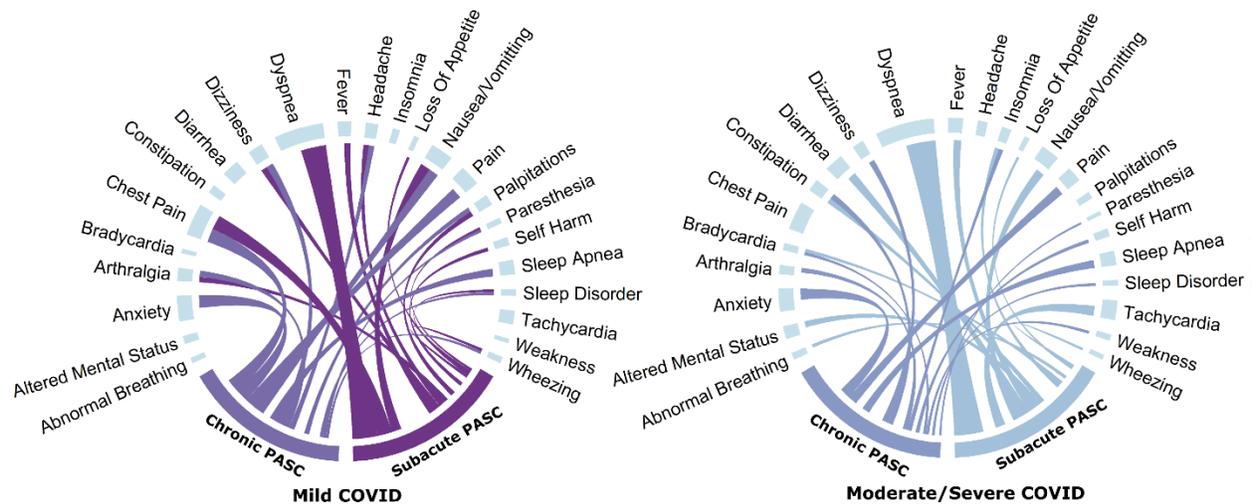

**Figure 3.** Chord plot illustrating symptom frequency of the most common symptoms for each of the four phenotypes. The twenty-four symptoms shown were selected from the union of the top twenty symptoms by frequency for each phenotype. Links are highlighted if the symptom frequency for that phenotype was higher than the average frequency across the four phenotypes. Links are also ordered counterclockwise from highest to lowest frequency for each phenotype.

**Discussion**
In this study, we developed a set of computable phenotypes for PASC using reproducible meta-heuristics based on COVID-19 severity and symptom duration. Although developing PASC phenotypes is useful for characterizing the complexity of the disease, several knowledge gaps remain. Notably, prior to this study, there were no studies that jointly evaluated COVID-19 severity and PASC symptom duration. We found several differences in symptom frequency between those with mild and moderate/severe COVID-19, including a higher frequency of certain gastrointestinal symptoms (such as constipation and diarrhea) in those with moderate/severe COVID-19 and a higher frequency of certain cardiopulmonary symptoms (such as chest pain and palpitations) in those with mild COVID-19. Additionally, patients with moderate/severe COVID-19 and subacute PASC had a higher frequency of cardiopulmonary symptoms when compared to patients from the other phenotypes.

*Computable Phenotypes for PASC*
An important contribution of this study is the development of computable phenotypes for PASC. Recent phenotyping methodologies for PASC have been highly varied between studies, making it difficult to establish standardized, reproducible phenotypes. Previous phenotyping studies differ with respect to how PASC is defined relative to COVID-19 and which symptoms are considered diagnostic. While a four-week period following acute COVID-19 is a consistent timeframe for defining PASC, the start point for the four-week gap can be vague. For example, one study defines PASC as the presence of symptoms between 1 to 6 months following the COVID-19 index although it is unclear whether the index is the diagnosis date or another date representing the end of acute COVID-19.[22] Another study using a three-week period following viral infection similarly does not provide an exact onset date for PASC relative to the infection period.[9]

Additionally, to our knowledge, few phenotyping studies accounted for multiple infections when considering symptom presence. In a study where multiple infections were considered, the first infection was selected for analysis with the remaining infection periods discarded, although it was unclear how observations aligned to subsequent infection periods were separated from those aligned to PASC.[8] Therefore, one contribution of this study is the development of meta-heuristics that clearly define when PASC-susceptible periods begin relative to one or more SARS-CoV-2 infections. Although the definition and timeframes of interest for PASC will likely evolve, our framework can easily be modified to incorporate new or different timeframes for analysis.

After applying our meta-heuristics, we identified four phenotypes of patients based on COVID-19 severity (mild vs. moderate/severe) and duration of symptoms (subacute vs. chronic). There were some similarities in the symptoms identified in these phenotypes versus general literature on PASC, including the high prevalence of cardiopulmonary symptoms among patients diagnosed with PASC[8–10,13,14] and the higher proportion of females with PASC symptoms after mild COVID-19 when compared to males.[10] We also found a higher frequency of neuropsychiatric symptoms in chronic PASC following both mild and moderate/severe COVID-19 when compared to the other physiologic groups. There is similar evidence from a study[13] that assessed different timeframes of symptom duration in that several neuropsychiatric symptoms were only present in the later PASC stage (from 6 – 9 months), although this study did not consider COVID-19 severity. Nevertheless, a growing body of evidence suggests many individuals with PASC may experience the persistence of neuropsychiatric symptoms.

*Reproducible PASC Symptom Profile*
We also developed a symptom profile for use in phenotyping and diagnosing PASC in the absence of the ICD-10 code. There is currently a lack of standardized definitions for PASC based on symptomatology. While some symptoms, such as dyspnea, appear consistently across PASC literature,[8,13,14,22,23] other symptoms, such as dementia, are only considered in a handful of studies.[13] Thus, when developing the symptom profile, we assessed symptoms associated with PASC independent of the concepts present in the N3C data by using clinician review to narrow PASCLex[18] concepts into a symptom profile. Therefore, our symptom profile can be adapted to any EHR data using OMOP CDM standards and is not dependent upon the concepts present in the N3C data. However, since applying our symptom profile only captured around three quarters of the patients clinically diagnosed with PASC, we suggest further review by additional experts to ensure all diagnostic symptoms are represented.

*Study Limitations*
There are some limitations to the phenotypes and meta-heuristics developed in the study. First, our phenotypes were developed under the assumption that it is possible to identify a COVID-19 index date for all patients. However, due to differences in testing and recording procedures throughout the pandemic, index COVID-19 dates may not be easy to identify. Furthermore, data associated with the COVID-19 index, such as the visit type, may vary between hospital systems. This lack of data collection was observed in the current analysis since up to 10% of COVID-19 index dates were classified as unknown severity due to a lack of data. Additionally, while it is possible to match each COVID-19 index date to the predominant variant using the limited data set, the COVID-19 variant is not included in the current phenotyping analysis. Future work includes expanding the definition of the phenotypes to characterize PASC following infection by different SARS-CoV-2 variants. Finally, the phenotypes rely on the recording of PASC symptoms for each patient at multiple timepoints, especially for the comparison between subacute and chronic PASC. However, the presence of symptoms can be more likely a function of how often an individual visits healthcare professionals and how different symptoms are recorded at each visit than the actual prevalence of these symptoms in individuals who had COVID-19. Therefore, while all PASC phenotyping is currently dependent upon symptom presence (due to a lack of gold standard), this method of identifying and phenotyping PASC patients introduces bias from variations in healthcare reporting.

**Conclusion**
We developed computable phenotypes and meta-heuristics for identifying PASC patients independent of the ICD-10 code U09.9. This framework can be adapted to other data sources to standardize phenotyping analyses and ensure rigor and reproducibility between studies. As part of this framework, we also developed an expert-reviewed symptom profile for PASC using a common data standard to further promote standardization of PASC phenotyping methods.

**Authorship Statement**
Authorship was determined using ICMJE recommendations.

**Acknowledgments**
The analyses described in this publication were conducted with data or tools accessed through the NCATS N3C Data Enclave covid.cd2h.org/enclave and supported by CD2H - The National COVID Cohort Collaborative (N3C) IDeA CTR Collaboration 3U24TR002306-04S2 NCATS U24 TR002306. This research was possible because of the patients whose information is included within the data from participating organizations (covid.cd2h.org/dtas) and the organizations and scientists (covid.cd2h.org/duas) who have contributed to the on-going development of this community resource.[24] The content is solely the responsibility of the authors and does not necessarily represent the official views of the National Institutes of Health or the N3C program. The N3C data transfer to NCATS is performed under a Johns Hopkins University Reliance Protocol # IRB00249128 or individual site agreements with NIH. The N3C Data Enclave is managed under the authority of the NIH.

Information can be found at https://ncats.nih.gov/n3c/resources.

We gratefully acknowledge the following consortial contributors to this work:
Christopher Chute.

We also gratefully acknowledge the following core contributors to N3C:
Adam B. Wilcox, Adam M. Lee, Alexis Graves, Alfred (Jerrod) Anzalone, Amin Manna, Amit Saha, Amy Olex, Andrea Zhou, Andrew E. Williams, Andrew Southerland, Andrew T. Girvin, Anita Walden, Anjali A. Sharathkumar, Benjamin Amor, Benjamin Bates, Brian Hendricks, Brijesh Patel, Caleb Alexander, Carolyn Bramante, Cavin Ward-Caviness, Charisse Madlock-Brown, Christine Suver, Christopher Chute, Christopher Dillon, Chunlei Wu, Clare Schmitt, Cliff Takemoto, Dan Housman, Davera Gabriel, David A. Eichmann, Diego Mazzotti, Don Brown, Eilis Boudreau, Elaine Hill, Elizabeth Zampino, Emily Carlson Marti, Emily R. Pfaff, Evan French, Farrukh M Koraishy, Federico Mariona, Fred Prior, George Sokos, Greg Martin, Harold Lehmann, Heidi Spratt, Hemalkumar Mehta, Hongfang Liu, Hythem Sidky, J.W. Awori Hayanga, Jami Pincavitch, Jaylyn Clark, Jeremy Richard Harper, Jessica Islam, Jin Ge, Joel Gagnier, Joel H. Saltz, Joel Saltz, Johanna Loomba, John Buse, Jomol Mathew, Joni L. Rutter, Julie A. McMurry, Justin Guinney, Justin Starren, Karen Crowley, Katie Rebecca Bradwell, Kellie M. Walters, Ken Wilkins, Kenneth R. Gersing, Kenrick Dwain Cato, Kimberly Murray, Kristin Kostka, Lavance Northington, Lee Allan Pyles, Leonie Misquitta, Lesley Cottrell, Lili Portilla, Mariam Deacy, Mark M. Bissell, Marshall Clark, Mary Emmett, Mary Morrison Saltz, Matvey B. Palchuk, Melissa A. Haendel, Meredith Adams, Meredith Temple-O'Connor, Michael G. Kurilla, Michele Morris, Nabeel Qureshi, Nasia Safdar, Nicole Garbarini, Noha Sharafeldin, Ofer Sadan, Patricia A. Francis, Penny Wung Burgoon, Peter Robinson, Philip R.O. Payne, Rafael Fuentes, Randeep Jawa, Rebecca Erwin-Cohen, Rena Patel, Richard A. Moffitt, Richard L. Zhu, Rishi Kamaleswaran, Robert Hurley, Robert T. Miller, Saiju Pyarajan, Sam G. Michael, Samuel Bozzette, Sandeep Mallipattu, Satyanarayana Vedula, Scott

Chapman, Shawn T. O'Neil, Soko Setoguchi, Stephanie S. Hong, Steve Johnson, Tellen D. Bennett, Tiffany Callahan, Umit Topaloglu, Usman Sheikh, Valery Gordon, Vignesh Subbian, Warren A. Kibbe, Wenndy Hernandez, Will Beasley, Will Cooper, William Hillegass, Xiaohan Tanner Zhang. Details of contributions are available at covid.cd2h.org/core-contributors.